# Encryption Mechanism And Resource Allocation Optimization Based On Edge Computing Environment

Ruan Yanjiao

College of Information Science and Engineering, Hunan University, 410082,ChangSha,China

ryj@hnu.edu.cn

**Abstract**.

A method for optimizing encryption mechanism and resource allocation based on edge computing environment is proposed. A local differential privacy algorithm based on a histogram algorithm is used to protect user information during task offloading, which allows accurate preservation of user contextual information while reducing interference with the playback decision. To efficiently offload tasks and improve offloading performance, a joint optimization algorithm for task offloading and resource allocation is proposed that optimizes overall latency. A balance will be found between privacy protection and task offloading accuracy. The impact of contextual data interference on task offloading decisions is minimized while ensuring a predefined level of privacy protection. In the concrete connected vehicle example, the method distributes tasks among roadside devices and neighboring vehicles with sufficient computational resources.

# 1 Introduction

In the edge computing environment, location-based services inevitably offload tasks to the end devices of the environment, how to reasonably make decisions about offloading tasks and allocate available computing resources is also a new challenge, existing offloading decision algorithms require contextual information about the mobile user's device, including location, speed and direction, etc. ), e.g., does not meet the needs of connected vehicle applications for timely reallocation of computational tasks to the cloud. Due to the increase in computing power of VANETs and vehicles, some connected vehicles (CVs) are now capable of solving complex computing tasks, and connected vehicles have become the new computing vehicles that access content and computing services from the Internet [1]. Vehicles as a new form of edge computing can reduce the latency of computing tasks by offloading the tasks to the edge devices. Scheduling and allocating computing resources to roadside units (RSUs) is essential for offloading tasks. The latency of offloading tasks is reduced by using contextual vehicle information to determine task support and resource allocation over time. However, some malicious attackers use this information to perform precision attacks that can result in vehicle collisions or even injuries. The data collected at the same time can be made available to third parties for data mining, but the user's privacy can be obtained, for example, by differential attacks on a dataset through multiple statistical queries, so in addition to avoid the violation of the user's privacy at the base station, the risk of privacy leakage of the statistical information published by the base station must also be considered [2]. The existing task offloading mechanism does not take into account the protection of the user's context-sensitive privacy. Therefore, it is necessary to investigate how to protect user privacy while efficiently transmitting computational tasks to the end-devices, i.e., to maximize the average total latency of the edge computations.

In recent years, the optimization of static edge computing scenarios has evolved, but only a limited set of load tasks can be performed within the static edge node coverage. However, currently many scenarios are dynamic, e.g., telematics network connections are dynamic and access and requests to servers change rapidly, which cannot keep up with the regular huge computational demands generated by urban intelligent transportation systems and information gatherers [3].

To address the above problems, this paper develops a method to optimize the encryption and resource allocation mechanism based on a edge computing environment. To protect privacy, the privacy differentiation technique is used in this paper to encode the user contextual information and add noise to meet ε-differentiated privacy requirements, but the contextual information after privacy-preserving processing affects the accuracy of task offloading decisions [3]. Therefore, the main problem is how to enable the decision center to make reasonable decisions about task offloading and resource allocation based on the encoded contextual information, and how to strike a balance between privacy protection and task offloading accuracy. By ensuring the desired level of privacy protection, the impact of corrupted contextual information on task offloading decisions is minimised. More specifically, this work proposes an optimized algorithm for protecting user contextual information such that contextual information, such as location and speed, satisfies ε-differentiated privacy protection for a given privacy budget ε. The decision center makes task offloading decisions based on contextual information processed with differentiated privacy protection, thus protecting user privacy, while the algorithm significantly reduces privacy protection. The algorithm significantly reduces the impact of privacy protection mechanisms on task offloading decisions and allows the decision center to make better offloading decisions. Then, this paper develops a task offloading decision method based on the domain space limiting algorithm to optimize the task execution delay by scheduling tasks on the server server and allocating resources to the server server to ensure the execution efficiency of the decision mechanism through optimization. Finally, this paper demonstrates the feasibility and efficiency of the two proposed methods through sufficient experiments, and the proposed method still provides high-quality solutions while protecting user privacy.

## 2 A Joint Optimization System Model For Privacy Protection And Resource Allocation

### 2.1 General overview

Due to the deployment of various privacy technologies, a key issue for resource allocation is how the decision center can make a reasonable decision on task offloading and resource allocation based on the breach context and how to balance the accuracy of task offloading and the protection of user privacy. Therefore, the system must minimize the impact of the disturbed contextual information on task reassignment decisions while ensuring that the necessary level of privacy protection is achieved. Through task scheduling and joint optimization, the best possible average latency can be achieved [4].

The structure of the resource allocation optimization scheme is as follows.

(1) The user does not request the required result from the local cache and uploads the task offloading request, and the DC in the client receives information about the task offloading request, including task information as well as the user's own location and contextual information.

(2) The DC uses the task offloading information collected from the user and the current server state information, such as the path objects, to analyze and execute the optimization decision-making algorithm and make decisions related to task offloadinging and resource allocation.

(3) The client forwards the user task to the designated server to execute the task, which is jointly executed by the local node and the cloud server.

In this scheme, in order to ensure the privacy and security of the user's data as much as possible, a partial offloading of the task is considered, i.e. the task is not fully executed between the edge server and the cloud server, which is a series of consecutive tasks issued by the user as one task and one computation of this task as a subtask, in which the computation process can be divided into two parts, i.e., the local computation part and the server computation part. In the case of issuing query tasks, which in most cases are designed to ensure user privacy, the risk of privacy breach can be appropriately mitigated by localizing the local computation part of the task [4].

**2.2 Problem Description**

For example, consider a roadside application scenario, in this scenario there is a base station B, roadside devices $R = \{R_1, R_2, \ldots, R_r\}$, all users in the roadside device $V = \{V_1, V_2, \ldots, V_v\}$, an edge server $M_i$ installed next to a user $V_i (i = 1, \ldots, v)$, an edge server $M_j$ installed next to a roadside device $R_j$, and $j = i + v$. The computational power of $M_i$ is $C_{M_i}$. When $M_i$ is near the Internet of cars $V_i$, $C_{M_i} = C_{V_i}$. And when $M_i$ is near ta roadside device $R_j$, $C_{M_i} = C_{R_j}$. $T_i$ represents a task of $V_i$, which can be divided into several sequential subtasks during task execution [5]. Each subtask may be executed jointly by a telematics server and an edge server, and the subtasks $T_i$ for each consecutive task contain the attributes $W_{T_i}$, $I_{T_i}$, $LO(T_i, \lambda_{T_i})$ and $EO(T_i, \lambda_{T_i})$, and we denote the total workload and the total amount of input for the subtasks by $W_{T_i}$, $I_{T_i}$, respectively. Also we denote the ratio between the output and input data of the locally executed task and the ratio between the output and input data of the task executed on the server by

$LO(T_i, \lambda_{T_i}), EO(T_i, \lambda_{T_i})$, respectively. The percentage of the workload replicated on the edge server $\lambda_{T_i}$. Different parameters are available for different computational tasks.

The data transmission channel between the CV and the edge server is typically a slow-frequency block diffuse channel, $V_i$ with the task $T_i$ and edge server $M_j$ at time $t_k$ the transport rate is $R_{t_k}^{V_i,M_j}$. The small-scale fading channel power gain between $V_i$ and $M_j$ is expressed as $h_{V_i,M_j}$, $\theta$ denotes the path loss index. $R_{t_k}^{V_i,M_j}$ can be calculated as follows [6].

$$R_{t_k}^{V_i,M_j} = B \log_2(1 + \frac{P_{M_j} h_{V_i,M_j} k_0 \|p(V_i,t_k) - p(M_j,t_k)\|^{-\theta}}{\alpha^2}) \tag{1}$$

$o(T_i, M_j, t_k)$ represents the offloading decision of the task $T_i$ to the edge server $M_j$ in time $t_k$, and $a(T_i, M_j, t_k)$ represents the resource allocation of the task $T_i$ to the edge server $M_j$ in time $t_k$. The delay of task execution consists of three parts: the delay of subtask execution on the CV $V_i$, and the transmission delay between $V_i$ and $M_j$, and the delay of subtask execution when the offloading decision $o(T_i, M_j, t_k) = 1$ is offloaded. $et(T_i, M_j)$ and $et(T_i)$ respectively represent the subtask $T_i$ of $M_j$ delay and the local delay. $ut(T_i, M_j) + ut(M_i, T_j)$ represents the delay of the data transmitting the task $T_i$ between the local CV and the edge server, $ut(T_i, M_j)$ represents the delay of the CV transmitting the data to the edge server and $ut(M_i, T_j)$ represents the delay of transferring the data from the edge server to the CV. The four parts are calculated as follows [6]:

$$et(T_i, M_j) = \frac{W_{T_i} \times \lambda_{T_i}}{a(T_i, M_j, t_k)} \tag{2}$$

$$et(T_i) = \frac{W_{T_i} \times (1 - \lambda_{T_i})}{C_{T_i}} \tag{3}$$

$$ut(T_i, M_j) = \frac{I_{T_i} \times LO(T_i, \lambda_{T_i})}{R_{t_k}^{T_i,M_j}} \tag{4}$$

$$ut(M_i, T_j) = \frac{I_{T_i} \times LO(T_i, \lambda_{T_i}) \times EO(T_i, \lambda_{T_i})}{R_{t_k}^{T_i,M_j}} \tag{5}$$

The optimization goal of the problem studied in this section is to obtain the decision after the optimization delay, which refers to the average task delay in the optimization system. Task $T_i$ delay:

$$De_{T_i} = et(T_i, M_j) + et(T_i) + ut(T_i, M_j) + ut(M_i, T_j) \tag{6}$$

The optimization goal is to minimize the average latency:

$$\min \frac{1}{n} \sum_{i=1}^{n} De_{T_i} = \min \frac{1}{n} \sum_{i=1}^{n} (et(T_i, M_j) + et(T_i) + ut(T_i, M_j) + ut(M_i, T_j)) \tag{7}$$

The edge server $M_i$ that requires the base station to allocate has less resources than the resources available, that is:

$$\sum_{M_i} a(T_i, M_j, t_k) \cdot a(T_i, M_j, t_k) \leq C_{M_j} \tag{8}$$

In time $t_k$, task $T_i$ can only be offloaded to one edge server, namely [7]:

$$\sum_{M_i} o(T_i, M_j, t_k) = 0 \text{ or } 1 \quad (9)$$

However, taking the total delay as the optimization target will lead to long tasks obtaining more resources, while short tasks obtaining insufficient resources. With the average reduction rate of system tasks as the optimization target, we can schedule tasks and allocate resources fairly for long and short tasks,

$$r(De_{T_i}) = \frac{De_{T_i} \cdot C_{T_i}}{W_{T_i}} \quad (10)$$

$r(De_{T_i})$ represents the latency reduction rate for all subtasks in task $T_i$, and the average task reduction rate:

$$\frac{1}{n}\sum_{i=1}^{n} r(De_{T_i}) \quad (11)$$

So the optimization objective is to minimize the average task reduction rate [8]:

$$\min \frac{1}{n}\sum_{i=1}^{n} r(De_{T_i}) \quad (12)$$

## 3 Method Introduction

### 3.1 Differential Privacy Algorithm Based On The Histogram

When a user offloads a request, their device will need to log into the Decision Center and report personal data, including location and speed, after being alerted by differentiated privacy technology. The Decision Center selects a specific server or cloud server to execute the request task sent by the user based on the base stations in the coverage area and allocates the corresponding resources using the branch bounding method to obtain the task reissue and resource allocation decision.

Differential privacy protects user privacy by adding controlled random noise to sensitive data. Initially, Dwork et al. proposed a concept of privacy in the literature based on a rigorous and provable mathematical definition of privacy, which provides a rigorous and provable notion of privacy. The content of a user's private information strictly corresponds to ε-differential privacy, ε-differential privacy is defined as follows.

**Definition** 4.3 (ε-differential privacy) [9]: Suppose a random algorithm A (where the algorithm is randomized means that its output is not a fixed value for a certain input, but follows a certain distribution) and a positive real number $\epsilon$. If for any adjacent dataset $DB$ and $DB'$ ($|DB - DB'| \leq 1$) that differ on a single element and any output $S \subseteq Range(A)$, the algorithm A provides ε-differential privacy,

$$P_r[A(DB) \in S] \leq e^\varepsilon \times P_r[A(DB') \in S] \quad (13)$$

In essence, differential privacy methods essentially transform an exact dataset query into a distributed query by determining the probability that two adjacent dataset queries return the same result. Differential privacy algorithms protect private data by mixing the information so that an attacker cannot distinguish between adjacent datasets and their results. The parameter ε determines the privacy protection capability, if ε is small, it means that DB and DB' are close, the more effective is the privacy protection provided by the differential algorithm. In the case of a randomized algorithm A, it is difficult to infer the existence of specific data from the results A(X) and A(Y) if the results for two adjacent data sets and are similar.

Differential privacy protection can be divided into local differential privacy protection, where each user can independently process their personal data before uploading it to a data warehouse, and centralized differential privacy protection techniques, where a trusted third party uses differential privacy algorithms to process statistics before publishing them. At this point, local differentiated privacy is used to encrypt the user's raw data, and the user's device transmits the encrypted data to the DC. This scheme does not assume that the DC is trustworthy, and the DC cannot accurately and in real time determine the user's location when the statistics are published or illegally accessed [9]. By applying differential privacy protection techniques to the user's contextual information, we can make the dataset provided by the DC decision center to third-party users meet the requirements for differential privacy protection, i.e., privacy protection can be proven while ensuring that the user's actual contextual information is not revealed in the event of a data leak or illegal access to the decision center.

The histogram-based algorithm can be used if the data collector is on top of the corresponding third-party query. As shown in Figure 1, the data collector can obtain an interference histogram using an efficient MWEM algorithm according to the query in Figure 1(b).

Different locations and speeds affect the user detection by the attacker differently. For example, if all vehicles are moving at the same speed, the attacker will not be able to detect the vehicle. However, if the vehicle is moving much faster or much slower than the others, the attacker can easily detect it. In this system architecture, differentiated privacy techniques are used when vehicles transmit contextual information. To reduce the errors associated with differential privacy, a local differential privacy algorithm based on random responses is proposed in this work.

In the differential privacy model, adding noise is easy by adding independent and homogeneously distributed Laplace noise to each unit lattice of the spatial histogram, while in reality choosing the lattice size is difficult in general practice: If it is too large, the accuracy of the query results obtained will not meet the requirements; if, on the other hand, the grid is too fine, most of the grid values will be sparse and the cumulative effect of the errors will be too strong, and if the cumulative effect is too significant and too wide, the accuracy of the results will drop sharply, which is a particularly acute problem when expressing large histograms [10].

To minimize the negative impact of differential privacy on the decision center, a new local differential privacy algorithm is proposed in this section together with the MWEM algorithm. The first task to be replicated is to obtain historical CV context statistics of connected vehicles, which include the distribution of different speeds and locations. For example, Figure 1 can be represented as a histogram of differential locations, each representing the cumulative number of vehicles per meter on the road. As shown in Figure 1(b), after running the histogram-based differential privacy algorithm , CV obtains a new histogram that it publishes to the decision center. Figures 1(a) and 1(b) show the histogram of the true statistic and the histogram after the differential privacy algorithm, respectively.

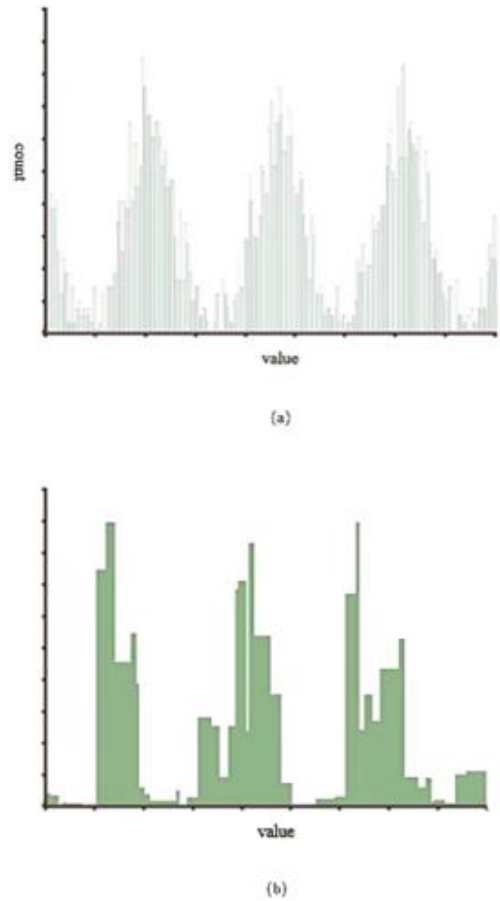

Figure 1 Histograms of raw data and data processed with MWEM

In the MWEM algorithm, reuse the exponential mechanism to find queries q beyond the current approximation to the underlying private data, measure these queries using the Laplace mechanism, and then use this measurement to improve the distributed multiplication weight update rules, assuming a set of queries exists $q \in Q$, and each function gets dataset A by $q(A) = \sum_{x \in D} q(x) \times A(x)$ going from the recording domain D to the interval [-1, + 1]. For dataset B sampling the query qi using $S_i(B,q) = |q(A_{i-1}) - q(B)|$ an exponential mechanism of ε parametrization was obtained. Noise is injected via the Laplace mechanism $m_i = q_i(B) + Lap(2T/\varepsilon)$, where T is the number of iterations. Let $A_i$ satisfy $A_i(x) \propto A_{i-1}(x) \times exp(q_i(x) \times (m_i - q_i(A_{i-1})) / 2n)$ n times of the distribution.

For any dataset B, the linear query Q, T ∈ N, ε> 0, MEWM will make A satisfied [11]

$$\max_{q \in Q} |q(A) - q(B)| \leq 2n\sqrt{\frac{\log|D|}{T}} + \frac{10T\log|Q|}{\varepsilon}$$

（14）

With this scheme HCs can obtain historical statistics for different regions or different periods. We obtain a histogram of position and velocity after a disturbance based on a differential privacy algorithm. It is assumed that the current velocity $v, v \in [v_1, v_2]$, $[v_1, v_2]$ is obtained from the algorithm and used to respond to the same values. CV then randomly selects one of the disturbed velocities $v' \in [v_1, v_2]$ as the decision center for reporting. CV then processes the location and a range $[x_1, x_2]$ is obtained for the statistics of the number of positions in the 2D coordinate system using MWEM and a randomly selected location is reported to the decision center.

The velocity v and location x vary but are usually within a limited range. So setting $v \in [0,100]$ $(m/s)$, but not a fixed range $x, x \in [0,1000]$ m is set in the histogram as the bearing position, i.e. the bearing position is set in a different grid and the 1000 m distribution is discussed. This configuration makes sense since an attacker cannot distinguish a specific user even at a distance of 1000 m if the location is spread out and there are multiple users along the path who are in range. Since the histogram obtained by the algorithm corresponds to differentiated privacy, the response of the decision center also corresponds to differentiated privacy when the base station sends the received information directly to a third party.

The user first needs to obtain vehicle statistics from the base station in order to offload the task in the region, which may be historical data or real-time statistics. This solution is not related to obtaining statistical information. The speed and location statistics are then transmitted via MWEM in a new histogram, and the end-user device encodes the speed and location from the last histogram and sends the encoded contextual information to the decision center [11].

**3.2 Task Offloading Decision Optimization Algorithm**

This subsection focuses on balancing offloading deployment efficiency and optimizing task offloading decisions. When a task is offloaded, it is first logged in the details reported by the decision center, and since offloading tasks involves sharing resources, the decision to offload that task may affect other tasks. In the local optimization approach, the decision center obtains relevant information about the task, including the user's location and speed at that time, the task's I/O connection from the user to the server, and the task's I/O connection from the server to the user. The decision center first finds the most appropriate task and then decides whether to offload it. In the initial task set, the decision center finds other m-1 tasks related to the initial task, after all m found tasks are marked as $O=\{T_i,...\}$, the decision center returns to execute the decision. Assuming that the other tasks are still in progress at the time the decision is made, the affected m tasks switch servers when the decision is made to offload the tasks, and the switching cost is not considered in this method. Specifically, the decision center receives the task registration information and first searches around to find available server servers when pending offload tasks are added to the candidate tasks until m candidate tasks are found or no more tasks are affected. The final offloaded tasks and the resource allocation policy are then obtained by a branch definition method.

The proposed branch bounding algorithm solves the server selection and resource allocation problem , which uses depth-first traversal to find an optimal solution. An asymptotic allocation method is used for the resource allocation, where the solution space and the mth problem are denoted by $s$ and $T_k$, respectively. In simple language, the user device CV of the playback problem owns the resources and in order to playback when the server server is selected, the branch allocation algorithm assigns a fixed resource C in the first resource allocation and assigns a resource 2C , and in the process of resource allocation, it must satisfy equation (8). Once the server resources are allocated, this must correspond to the edge value of the current solution being smaller than the minimum value of the existing solutions, i.e., the optimal solution can be found from the current choice s. The algorithm then updates the current solution s and moves the solution s to the global traversal stack. If all tasks in the solution space s have completed server selection and resource allocation, and the delay is less than the current minimum, the algorithm assigns the current optimal solution to the solution space s until the end of the traversal. If the global stack is empty, the branch-and-bound algorithm returns the optimal solution [12].

If the solution for m tasks has already completed resource allocation, the edge value is only relevant for m-1 tasks. To obtain the edge value, we randomly select the edge servers and allocate the maximum resources without considering the data transfer time between the edge server and the user, and finally compute the solution delay s. In this case, the latency of the task is used as the boundary value to achieve the tuning effect.

## 4 Experiment

### 4.1 Experimental Setting

The experiment is designed for a one-way multi-lane road of size 2000 m * 2000 m, with each lane being 3 m wide. The roadside devices, which can act as servers, are located 20 m from the far end of the road and 10 m above the road. The transmission power of each CV and roadside device is set to 100 mW and 10 mW, respectively, to test the performance of the system at the time of deciding to offload the task.

The parameters of the simulation tests are given as follows. The roadside unit $R_i$ includes its location $p(R_i)$ and the capacity of the processing task $C_{R_i}$. The tasks include their initial location $p(V_i, t_0)$, initial speed $v(V_i, t_0)$, the computational power of vehicle i and the total workload of the subtasks $W_{T_i}$, the total amount of input data of the subtasks $I_{T_i}$, the ratio between the output and input data of the tasks running locally $LO(T_i, \lambda_{T_i})$, and the ratio between the output and input data of the tasks running on the server $EO(T_i, \lambda_{T_i})$. Indicates how much of the workload is replicated on the server side. Vi of a vehicle that can act as an edge server includes its location $p(V_i, t_0)$ and its ability to execute tasks $C_{V_i}$.

In this paper, the following three task offloading methods are selected as reference algorithms, and the joint optimization of resource allocation and task offloading proposed in this chapter is compared with the following reference algorithms.

**RM**: Randomly selected maximum resource allocation algorithm, in which the algorithm randomly selects the edge server that is most capable of further connectivity, and then allocates all resources on the edge server to the task to obtain the solution result [12].

**CM**: A nearest neighbor server selection and maximum resource allocation algorithm, where the algorithm selects the nearest neighbor server and then allocates all the resources on the nearest neighbor server to the task to obtain the solution result.

**BM**: Edge server selection based on branch limiting method and maximum resource allocation algorithm, in which the algorithm selects an edge server based on the branch limiting method and then allocates all the resources of the edge server to the problem to obtain the solution result.

For this experiment, there are two main evaluation metrics.

1) Mean latency reduction. Using the average latency reduction rate as the final latency optimization objective. The latency reduction ratio is the estimated ratio between the number of tasks that are offloaded and the number of tasks that are not offloaded; equation (10) determines the latency reduction ratio, and equation (12) determines the average latency reduction ratio.

2) Task multiplier. To evaluate the throughput of the proposed task offloading algorithm, a task multiplier is introduced. Since task offloading uses partial repetition, where the server and the user device process the task simultaneously, the proposed computational model can improve the task throughput. This is similar to the process in a pipeline, where the total number of consecutive subtasks per unit time is used as a criterion. The task throughput per unit 2 is the ratio between the number of tasks completed with playback and without playback.

In this section, three experiment s are planned to verify the feasibility of this solution.

Experiment 1, examines the impact of differential privacy on the estimation of the average delay reduction rate of metric 1 and the task 2 multiplier using the playback decision algorithm presented in this section. The effectiveness of the RR algorithm and our proposed privacy-preserving algorithm was tested by observing the changes in the evaluation metrics for different privacy-preserving algorithms using different privacy-preserving principles, while empty control experiments were conducted, i.e., no privacy-preserving algorithm was applied and the user location and speed data were uploaded directly.

Experiment 2 verifies the usefulness of the offloading decision algorithm presented in this chapter and three other decision algorithms RM, CM and BM in evaluating the metric. The amount of latency reduction and task throughput of our proposed offloading decision algorithm is compared with the other three decision algorithms.

Experiment 3 confirms the effect of the offloading decision algorithm in this section on the performance of two different privacy algorithms on two evaluation metrics with different privacy budgets ε. The effects of the RR algorithm and our proposed privacy-preserving algorithm on the two evaluation metrics were tested for privacy budgets ε of 1, 5, 10, and 20, respectively.

**4.2 Experimental results and analysis**

In the experiments, different algorithms were used to offload all the baseline tasks. Then, the experimental evaluation metrics 1 and 2 were obtained by a simulation process as shown in Figures 2-9, where the horizontal coordinates represent the timing of the simulation process, and the other half of the horizontal coordinates keeps the trend of metrics 1 and 2 constant after all the tasks are executed, i.e., our reproduction solutions and privacy-preserving algorithms were validated

In Experiment 1, we compare the results of our proposed task offloading solution under three privacy-preserving models. The performance metrics are the average delay reduction and the task replicator, and in this experiment the privacy budget ε is set to 5 for both RR and our own privacy-preserving algorithm, as shown in Figures 2 and 3, where a represents the use of our decision algorithm alone without privacy-preserving algorithm; b represents the use of our proposed decision algorithm and privacy-preserving algorithm RR; and c represents the use of our proposed decision algorithm and privacy-preserving algorithm RR. privacy-preserving algorithm. When the differentiated privacy-preserving algorithm is not used, our decision algorithm achieves a relatively small average delay reduction, since the decision center can directly report the exact location and rate. When the RR algorithm is used, the average latency reduction is relatively large. When using our privacy-preserving algorithm, the average delay of our algorithm is less than that of the RR algorithm. The comparative results of our decision algorithm for offloading tasks on evaluation measure 2 under three privacy models are shown in Figure 3. It can be seen that our privacy-preserving algorithm also reduces the impact of the privacy budget ε on the experimental evaluation measure 2 compared to the benchmark RR algorithm.

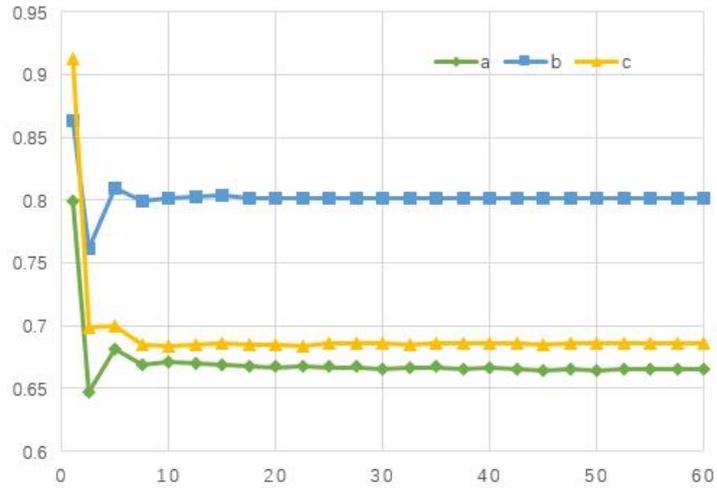

Figure 2 Effect of decision algorithms on evaluation metric 1 under three privacy protection modes

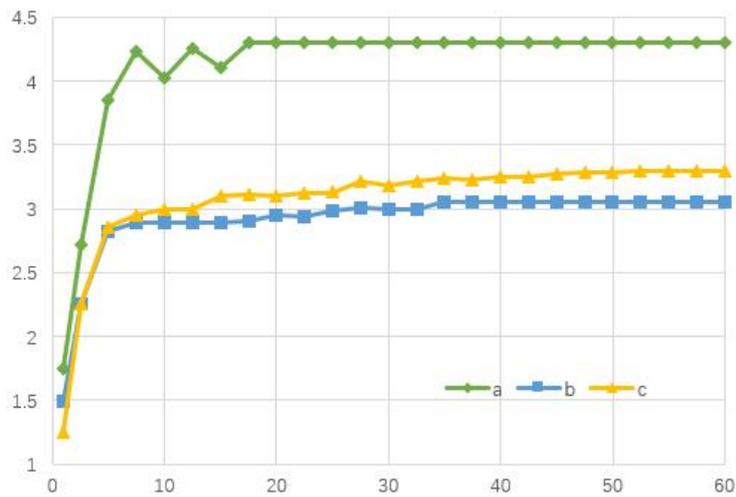

Figure 3　Effect of decision algorithms on evaluation metric 2 under three privacy protection modes

In Experiment 2, we compared our proposed problem with the RM, CM, and BM offload algorithms. Figure 4 shows the simulation results of evaluation measure 1 for the four algorithms. The a represents the RM algorithm, b the CM algorithm, c the BM algorithm, and d our proposed decision algorithm. Among them, our decision offloading algorithm has the smallest average delay reduction, which proves that our algorithm can effectively reduce the overall delay. Figure 5 compares the simulation results of the four algorithms for evaluation method 2. It can be seen that our proposed method has the maximum task multiplier, which proves that our algorithm can achieve higher task throughput.

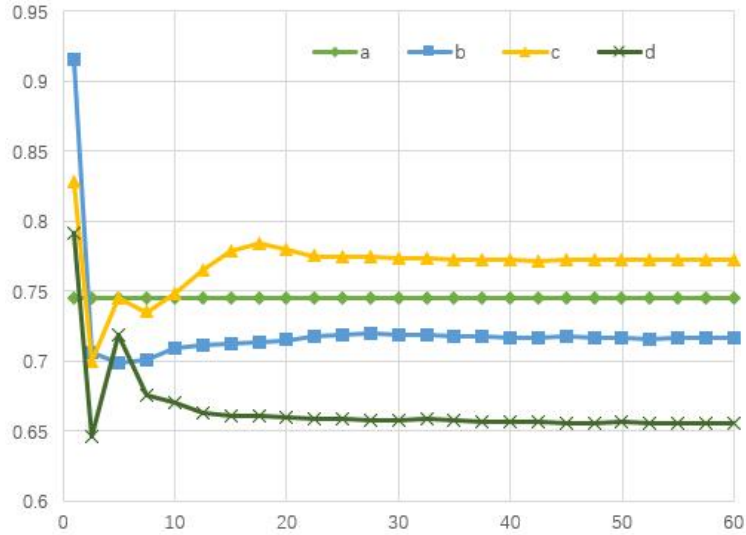

Figure 4 The influence of the different task-offloaded decision algorithms on the evaluation index 1

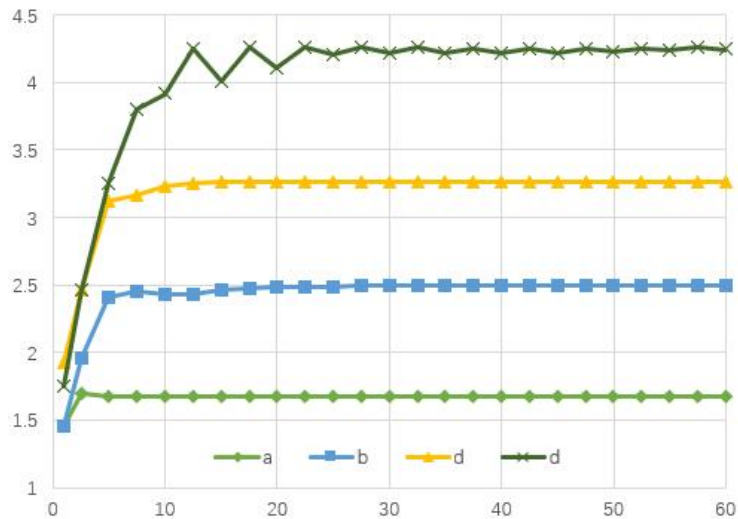

Figure 5 The influence of the different task-offloaded decision algorithms on the evaluation index 2

In Experiment 3, we compare the impact of RR and our privacy-preserving algorithm on our task offloading decision algorithm when the privacy budget ε is variable. Figure 6 and Figure 7 show the simulation results obtained with the RR algorithm for different privacy budgets ε. It can be seen that as ε decreases, the average delay decreases and the number of completed tasks decreases. The final curves for ε = 10 and ε = 20 overlap as contextual information is included with less noise. It can be concluded that the error of RR is larger than the error of our proposed decision algorithm, which leads to inaccurate results in the decision to offload the task. Figure 8 and Figure 9 show the simulation results obtained with our privacy-preserving algorithm under different privacy budgets ε. It can be seen that these lines are very close, i.e., the violation of the user's contextual information by our privacy-preserving algorithm has less impact on the task offloading decision. The small impact of the privacy budget on the offload algorithm reflects the performance of the algorithm even when the privacy budget changes.

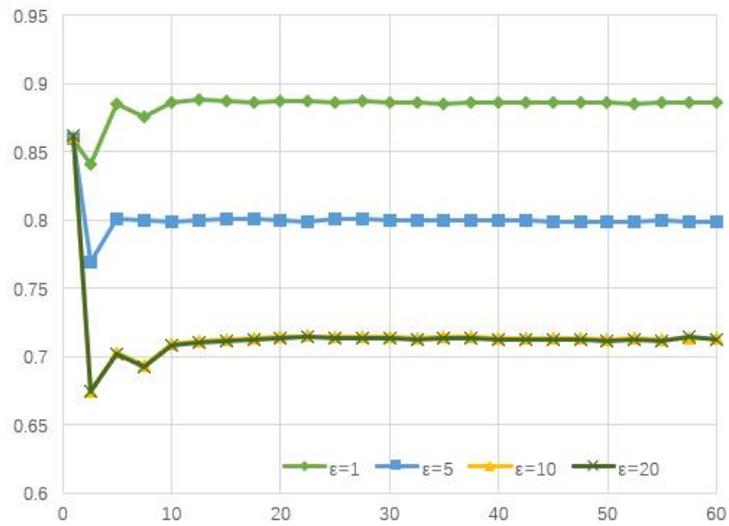

Figure 6 Effect of different privacy budget ε on evaluation metric 1 in the RR algorithm

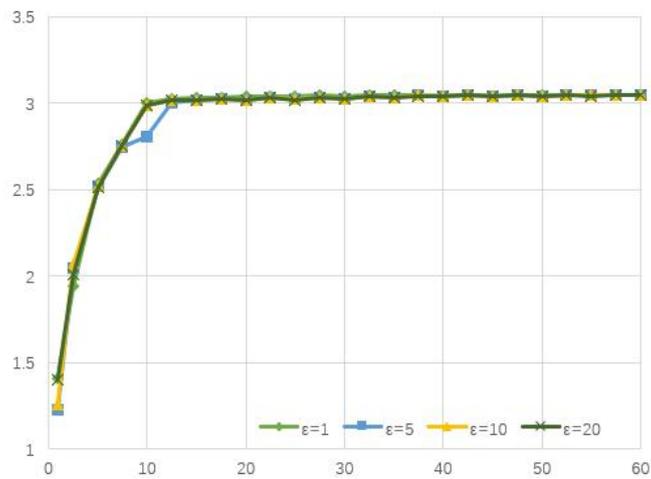

Figure 7 In the RR algorithm, the effect of different privacy budgets ε on the evaluation indicator 2

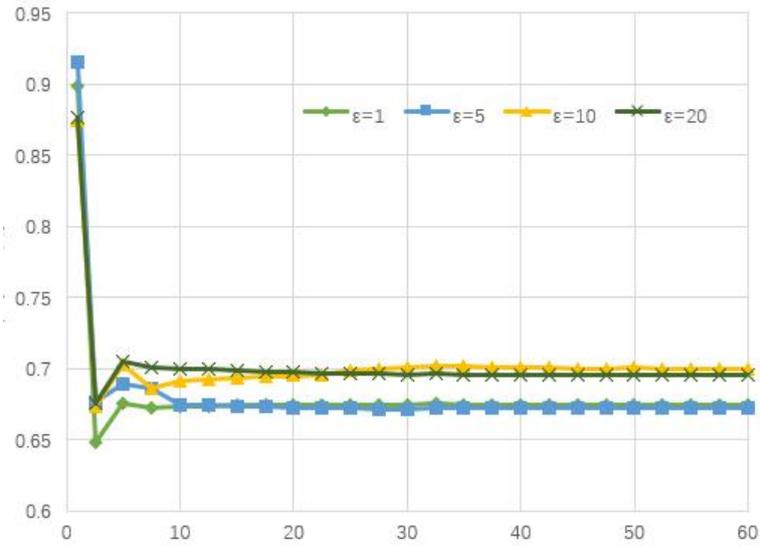

Figure 8 The impact of different privacy budget ε on the evaluation indicator 1 in the privacy protection algorithm proposed in this chapter

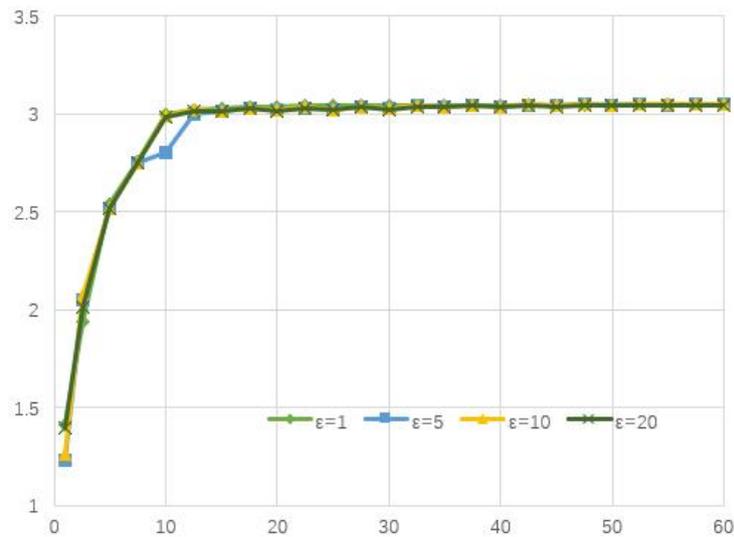

Figure 9 Impact of different privacy budgets ε on evaluation indicator 2 in the privacy protection algorithm proposed in this chapter

## 5 Conclusion

In order to protect the user's contextual information during task offloading, this paper presents a task offloading and resource allocation algorithm that includes a differentiated privacy algorithm. This method reduces its impact on the offloading decision while preserving more accurate contextual information for the user. To improve offloading performance, the paper also employs a domain bounding algorithm to optimize the scheme's latency. To verify the feasibility and effectiveness of this scheme, the method is evaluated through simulation experiments. The experimental results show that the privacy-preserving and branch-limiting algorithms proposed in this paper can effectively optimize the latency while guaranteeing the user's privacy and achieving higher task throughput, and reduce the impact of the privacy budget on the latency while protecting the privacy of the user's contextual information.